\documentclass[ 5p, 11pt, twocolumn, times]{elsarticle}
\biboptions{sort&compress}

\usepackage{amsmath}
\usepackage{siunitx}
\usepackage{lineno,hyperref}
\modulolinenumbers[5]

\newcommand*\diff{\mathop{}\!\mathrm{d}}
\usepackage{upgreek}
\usepackage{graphicx}
\usepackage{subfigure}
\usepackage{color}
\usepackage{soul,xcolor}
\setstcolor{red}

\DeclareMathAlphabet{\Ibb}{U}{msb}{m}{n}
\newcommand   {\IC}{{\ensuremath{\Ibb C}}}
\newcommand \ggp {{\ensuremath{{{\gamma}/{\gamma}^{\prime}}}}}
\newcommand \gp  {{\ensuremath{{\gamma}^{\prime}}}}
\newcommand \g   {{\ensuremath{\gamma}}}
\newcommand{\matone}{\ensuremath{\text{\textup{\textbf{I}}}}}

\newcommand{\revision}{}

\begin{document}

\begin{frontmatter}


\title{Insights from a minimal model of dislocation-assisted rafting in single crystal Nickel-based superalloys}

\author[mymainaddress]{Ronghai Wu}

\author[mymainaddress]{Stefan Sandfeld
\corref{mycorrespondingauthor}}
\cortext[mycorrespondingauthor]{Corresponding author}
\ead{stefan.sandfeld@fau.de}

\address[mymainaddress]{Institute of Materials Simulation, Department of Materials Science, Friedrich-Alexander University of Erlangen-N\"urnberg (FAU), Dr.-Mack-Str. 77, 90762
F\"urth, Germany}

\begin{abstract}
Nickel-based superalloys play a major role in many technologically relevant high temperature applications. Understanding and predicting the evolution of the phase microstructure during high temperature creep together with the evolution of the dislocation microstructure is a challenge that up to date has not yet been fully accomplished. Our two-dimensional coupled phase-field/continuum dislocation dynamics model explains  microstructural mechanisms which are important during the early stage of rafting in a single crystal system. It shows how \ggp\ phases and dislocations interact giving rise to realistic creep behavior; no phenomenological fit parameters are required.
\end{abstract}

\begin{keyword}
superalloy; rafting; phase-field model; continuum dislocation dynamics; microstructure
\end{keyword}

\end{frontmatter}

Superalloy development has been mainly driven by improved control of the phase microstructure, which in Ni-base superalloys essentially consists of cube-shaped precipitates of an ordered phase (termed $\gamma^\prime$ phase) embedded in a face-centred cubic, solution hardened matrix (termed $\gamma$ phase).  The L$1_{2}$ structure of the precipitates renders them strong obstacles to dislocation motion even at elevated temperatures, and optimizing the precipitate microstructure has played a major role for improving the thermo-mechanical properties of single crystal Ni-based superalloys. Due to their remarkable resistance to thermo-mechanical and chemical degradation in extreme service environments this class of materials became a common choice for components which require good creep resistance at high-temperatures \citep{Pollock1992_AMM40}.
One example for the demanding service conditions are turbine blades in 
jet engines where blade temperatures can be up to ${\approx}1000^\circ\textrm{C}$. Additionally, centrifugal forces in the blades cause sustained tensile stresses, giving rise to creep deformation at high-temperatures \citep{Mughrabi2000_AEM2, Reed2009_ActaMater57}. A problematic aspect for the life time of superalloy-based components under these conditions is rafting (directional coarsening) of the precipitate microstructure which leads to degradation of material properties. Understanding the  mechanisms which control the coupled evolution of precipitate and dislocation microstructure under stress, and understanding how this evolution controls the macroscopic creep behavior, is therefore an important step towards improving superalloys. 

Experimentally exploring the effect of process parameters, temperature and loading conditions, and at the same time collecting detailed information about the phase and dislocation microstructure and the mechanical response, is a complex task. Identifying and understanding all mechanisms that are responsible for the microstructure-property relationship is even more challenging and is until today not fully accomplished. As a naturally arising question one therefore may ask: is there a predictive modeling approach which can simultaneously provide information about the $\gamma/\gamma'$ evolution, the flow and interactions of dislocations, and the macroscopic creep properties? The most promising approach may be to couple two different methods to describe the stress-driven evolution of the dislocation microstructure and the concomitant plastic deformation on the one hand, and the phase microstructure ($\gamma/\gamma^\prime$) on the other hand. 

The phase-field (PF) method has become the most successful mesoscale simulation approach for predicting the $\gamma/\gamma'$ pattern evolution (see e.g. \citep{Steinbach2009_MSMSE17,Emmerich2008_AiP57}). 
\revision{Simulations which couple phenomenological constitutive material models (such as viscoplasticity) and the PF method show that plastic activity accelerates rafting and allows misalignments of rafts with respect to the cubic direction \citep{JMPS_2012_Cottura, PM_2010_Gaubert}. However, the flow and interaction of dislocations -- which are the fundamental carriers of plastic deformation -- can not be considered in those models. 
Pioneering steps towards using PF methods for representing dislocations have already been undertaken from the early 2000s on \citep{2000_MRS_Finel, 2001_Acta_Khachaturyan,2003_Acta_Finel,PM_2010_Zhou} and have been applied to a number of two- and three-dimensional situations including, e.g., the evolution of dislocation loops from a Frank-Read source, the annihilation between two attracting dislocation segments or also rafting of $\gamma/\gamma'$ alloys.}
\revision{The dilemma that these models face is twofold: 
	Most of these PF descriptions of dislocations have to use extremely small grid spacing, i.e., smaller than the magnitude of the Burgers vector \citep{2000_MRS_Finel} or smaller than the average dislocation spacing \citep{PM_2010_Zhou}. Additionally, for increasing the system sizes, simple spatial coarse-graining of systems of dislocations lead to a loss of short-range interactions between dislocations. This issue is still not fully resolved; steps towards an up-scaling suitable for large scale simulations were, e.g., introduced by Finel et al. \citep{2003_Acta_Finel}, allowing for dislocation core sizes smaller than the grid spacing while still retaining short range interactions between dislocations.}
\revision{The other problematic point is that order parameters associated with the plastic strain of a glide system, contain only information of the geometrically necessary dislocations (i.e., through plastic strain gradient), despite the fact, that the background of statistically stored dislocations still may contribute to plastic deformation. }

In the present work, we show an alternative approach by coupling a PF model and a 2D model of continuum dislocation dynamics (CDD) \cite{Acta_2003_Michael,MSMSE_2013_Stefan}. The PF model governs the $\gamma/\gamma^\prime$ evolution, while the CDD model is used to represent fluxes of positive and negative edge dislocations, from which geometrically necessary dislocations (GNDs) and statistically stored dislocations (SSDs) can be computed. 
\revision{One of the advantages of this model is that short-range interaction terms have already been systematically derived in \cite{Acta_2003_Michael}, which compensates for the loss of information due to coarse-graining.}
By superposition of dislocation-associated and phase-associated eigenstrains one can in a straightforward manner formulate the elastic energy which accounts for dislocation stresses, coherency stresses associated with $\gamma/\gamma^\prime$ misfit strain, and the mutual interaction between $\gamma/\gamma^\prime$ and dislocations. 

In the following we consider a plane strain geometry with periodic boundary conditions and one \gp\ precipitate in the center with an initially quadratic cross section. \g\ channels are populated by edge dislocations on two different slip systems, $i=\{1,2\}$,  which are characterized by  Burgers vector $\mathbf{b}^i$ and slip plane normal $\mathbf{n}^i$ (see Fig.1a). We consider a Ni-Al binary system with no distinction of $\gamma^\prime$ variants. A simple phase field model with a composition field $c$ governed by the Cahn-Hilliard equation is sufficient in this case, where $c=\text{c}_{\gamma}$ represents the $\gamma$ phase and $c=\text{c}_{\gamma^\prime}$ the $\gamma^\prime$ phase, respectively. The total energy functional
\begin{equation}
F = \int_\text{V} (f^{\text{bulk}} + f^{\text{grad}} + f^{\text{el}}) \diff \text{V}
\end{equation}
results from the bulk energy density
$f^{\text{bulk}} = \text{f}_0 (c-\text{c}_{\gamma^\prime})^2 (c-\text{c}_{\gamma})^2$,
the gradient energy density $f^{\text{grad}} = \frac{\text{k}_c}{2} |\nabla c|^2$ and the elastic strain energy density
$f^{\text{el}} = \frac{1}{2} \boldsymbol \sigma : \boldsymbol \epsilon^{\text{el}}$.
%
Therein, $\text{f}_0$ is the energy density scale and $\text{k}_c$ is the gradient energy density coefficient determined by fitting the calculated interface energy to an experimentally obtained interface energy. The stress tensor $\boldsymbol \sigma$ results from fulfilling mechanical equilibrium, $\nabla \cdot \boldsymbol \sigma = \boldsymbol 0$, in the absence of body forces. The stiffness tensor $\IC$ links elastic strains and stresses through the constitutive equation $\boldsymbol \sigma = \IC : \boldsymbol \epsilon^{\text{el}}$ and is in general different for $\gamma$ and $\gamma'$. However, the goal of this work is not an utmost realistic simulation but rather a minimal model with which relevant mechanisms can be identified. Elastic inhomogeneities due to different stiffness tensors as well as plastic inhomogeneities from dislocations both would contribute to rafting. To identify the influence of the latter we will therefore completely neglect different elastic properties and assume the same stiffness tensor for the \g\ as well as for the \gp\ phase.
In a small strain context one can additively decompose the total strain $\boldsymbol \epsilon$ into an elastic elastic strain $\boldsymbol \epsilon^{\text{el}}$ and two inelastic contributions, which are: 
(i) the \ggp\ misfit strain $\boldsymbol\epsilon^{\text{mis}} = \bar\epsilon^{\text{mis}} \alpha_c^3 (10 - 15\alpha_c + 6\alpha_c^2) \matone$ with $\alpha_c=(c-\text{c}_{\gamma})/(c_{\gamma^\prime}-\text{c}_{\gamma})$ and  the identity tensor $\matone$, and 
(ii) the dislocation eigenstrains $\boldsymbol \epsilon^{\text{dis}}$ which are obtained from the respective strains $\eta^i$ in each slip system $i$ by  $\boldsymbol \epsilon^{\text{dis}} = \sum_i \eta^i \mathcal{M}^{i}$. There, the transformation of dislocation strains $\eta^i$ from the local system into the global coordinate system is done by the projection tensor $\mathcal{M}^i = \frac{1}{2\text{b}^i}(\mathbf{b}^i \otimes \mathbf{n}^i + \mathbf{b}^i \otimes \mathbf{n}^i)$. \revision{The smooth fifth order polynomial interpolation for the $\ggp$ misfit is chosen instead of a linear interpolation function (following Vegard's law) for numerical reasons.}
%

The $\gamma/\gamma^\prime$ evolution is governed by the Cahn-Hilliard equation, which, for a homogeneous and isotropic interface mobility coefficient $\text{M}_c$, is given by
\begin{equation}
\label{eq:Allen-Cahn}
\frac{\partial c}{\partial t} = \text{M}_c \nabla^2 \frac{\delta F}{\delta c}.
\end{equation}
The dislocation microstructure evolution for each slip system is governed by a continuum dislocation dynamics model in which positive ($\rho^{+}$) and negative ($\rho^{-}$) edge dislocation densities can be distinguished \citep{Acta_2003_Michael}. Within this model one can easily obtain the total density $\rho = \rho^{+} + \rho^{-}$ or the excess (i.e signed GND) density $\kappa = \rho^{+} - \rho^{-}$. The evolution equations for $\rho^{+}$, $\rho^{-}$ and the plastic strain $\eta$ are given by
\begin{equation}
\label{eq:dislocation velocity law}
{\partial_t \rho^{+}} = -\partial_{x} (v \rho^{+}), \quad {\partial_t \rho^{-}} = \partial_{x} (v \rho^{-}), \quad {\partial_t \eta} = \rho v \text{b},
\end{equation}
where $x$ denotes the local coordinate in glide directions and $v$ is the dislocation velocity (the superscript $i$ indicating the slip system was dropped for brevity). Assuming a linear relationship between stresses and dislocation velocity, one can write \revision{
\begin{equation}
\label{eq:velocity stress}
v = \left
\{ 
\begin{array}{ll}
\frac{\text{b}}{\text{B}}\text{sign}(\tau)(|\tau| - \tau^{\text{y}}) & \text{if}\;\; |\tau| > \tau^{\text{y}},\\
0 & \textrm{else}
\end{array} 
\right.
\end{equation} }
where $\text{B}$ is the drag coefficient \revision{and $\tau=\tau^\text{l}+\tau^\text{b}$.} $\tau^\text{l}=\boldsymbol\sigma:\mathcal{M}$ is the long range shear stress resulting from external mechanical loading, heterogeneous plastic strain (i.e. dislocation eigenstrain) as well as from \ggp\ misfit. 
$\tau^{\text{b}}=-\text{DGb} {\partial_{x} \kappa}/{\rho}$ is the back stress governing the \revision{short-range} repulsion of like-oriented dislocations within a slip plane with the dimensionless parameter (\cite{Acta_2003_Michael}) $\text{D}=0.6$   and $\text{G}$ the shear modulus. 
$\tau^{\text{y}} = [({c_{\gamma^\prime}-\text{c}_{\gamma}})/({\text{c}_{\gamma^\prime}-c})] a \text{Gb} \sqrt{\rho^1+\rho^2} $ is the yield stress with $a=0.4$. Since here the early stage of high temperature/low stress creep is considered,  dislocations hardly cut through the $\gamma^\prime$ phase, which is reflected by the bracketed factor in $\tau^{\text{y}}$: inside the precipitate $\tau^{{\rm y}}\rightarrow\infty$ results in zero velocity, while outside the precipitate $\tau^{{\rm y}}$ is just the commonly used Taylor-type yield stress, with a smooth transition in between.

In our simulation the elastic eigenstrain problem is solved by a finite element method, the PF and CDD evolution equations are solved by the finite volume method. The quadratic simulation domain is periodic in both direction and has a size of $512 \times 512$ \SI{}{nm}; the mesh for all numerical methods consists of $64 \times 64$ quadratic elements. Since the velocity of \gp\ evolution is much slower than that of dislocation flow,  for each PF time step the CDD problem is solved with a number of smaller sub-time steps until a quasi-stationary dislocation configuration is reached. As initial condition, dislocations of both signs (represented by Gaussian density distributions) are randomly distributed in the $\gamma$ channels resulting in an average density of
$\bar\rho= 6 \times 10^{12}$\,\SI{}{m^{-2}} for each slip system, which is in the same order of magnitude as experimentally measured data \cite{MATEC_Kondo_2014, Acta_2013_come}. 
%
The initial \gp\ contour is shown as dashed line in Fig.1b. 
Material parameters correspond to a temperature of $1253\ldots\SI{1293}{K}$ with $\bar\upepsilon^{\text{mis}}$=$-0.003$, C$_{11}$=\SI{198}{GPa}, C$_{12}$=\SI{138}{GPa}, C$_{44}$=\SI{97}{GPa}, c$_\gamma$=$0.160$, $b$=\SI{0.25}{nm}, $B=10^{-13}$\SI{}{GPa s}, c$_{\gamma^\prime}$=$0.229$, M$_c$ =$5 \times 10^{-17}$ \SI{}{J^{-1}mol^{2}m^{-1}s^{-1}}, k$_c$=$5 \times 10^{-7}$ \SI{}{Jm^{-1}} \revision{and f$_0 = 1 \times 10^{12}$ \SI{}{Jm^{-3}}} \cite{Zhou_2008_Acta, Zhou_2007_Acta}. \revision{The calculated interface energy is \SI{0.036}{Jm^{-2}}; the interface region is then discretized by $6$ grids points.}
\begin{figure}[htp] \centering
	\hbox{}\hfill
	\includegraphics[width=0.85\columnwidth]{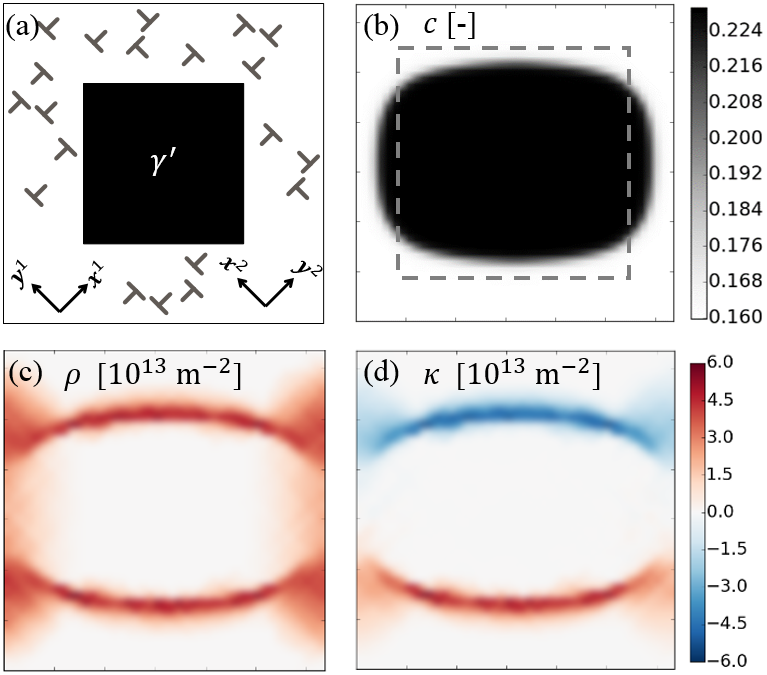}
	\hfill\hbox{}
	\caption{\label{fig:fig1_v1}
		(a) Schematic of the system (not to scale); (b) initial $\gamma/\gamma^\prime$ morphology (dashed line) and at $t=5.4\times 10^4$ \SI{}{s}; (c) total and (d) excess dislocation density at $t=5.4\times 10^4$ \SI{}{s}.	}
\end{figure}

The simulated system is subjected to an applied tensile stress \revision{of} \SI{200}{MPa} in the vertical direction. After a creep time of $t=5.4 \times 10^{4}$\,\SI{}{s}, the $\gamma^\prime$ precipitate is rafted in  horizontal direction with an aspect ratio of $1.35$ (Fig.1b). This {\lq}N-type rafting{\rq} (i.e. the rafting direction is normal to the external loading direction), is commonly observed in Nickel-based superalloys with negative $\gamma/\gamma^\prime$ misfit. At this stage the precipitate still has a roughly rectangular shape, the strip-like morphology has not occurred yet, in agreement with experimental observation for the early stage of rafting at high temperature and low stress \cite{ISIJ_2011_Miura}. The corresponding total and excess dislocation densities are shown in Fig.1c,d. It can be seen that dislocations predominantly accumulate at the horizontal $\gamma/\gamma^\prime$ interfaces. The \gp\ "polarizes" the dislocation structure such that positive edge dislocations are blocked at the lower horizontal interface, while negative dislocations accumulate at the upper interface. However, in the vertical channels no such GND accumulation occurs. Comparing total and GND density in the vertical channels one finds that there dislocations must be SSDs.
For this creep regime, the interface dislocation density typically increases from about $1 \times 10^{13}$ \SI{}{m^{-2}} up to $3\ldots 30 \times 10^{13}$ \SI{}{m^{-2}} \cite{MATEC_Kondo_2014, Acta_2013_come} (accurate measurement of interface dislocation densities is difficult and the {\lq}thickness{\rq} of the interface layer is not well defined). The simulated interface density of around $5 \times 10^{13}$ \SI{}{m^{-2}} is in good agreement with the experimental data.

In order to reveal the role of dislocations during rafting, we do simulations with and without dislocations and compare them at an intermediate time step, $t=2 \times 10^{4}$ \SI{}{s}. 
Fig.2c shows the long range stress, which in the case without CDD, consists of contributions from applied stress and the misfit. There, shear stresses in the horizontal channels are significantly higher than those in the vertical channels. Thus, dislocation activity (if present) would be suppressed in the latter. Fig.2a shows the evolved system with dislocations: in comparison to the system without CDD it can be seen that the maximum long range shear stress is reduced. The reason are dislocation pileups which develop as follows:
$\tau^l$ is positive in the horizontal channels and negative in the vertical channels. According to \eqref{eq:dislocation velocity law} and \eqref{eq:velocity stress}, positive stress drives positive dislocations into positive direction (accumulating at the lower horizontal interface), while negative dislocations move into negative direction (accumulating at the upper horizontal interface). This interface dislocation structure alters the stress field due to a combination of backstress (from pile-ups) and the mean field stress (from heterogeneous plastic strains). 
Ultimately, dislocation activity then also changes the elastic energy potential, which determines the evolution  of the $\gamma/\gamma^\prime$ composition field in \eqref{eq:Allen-Cahn}. In Fig.2b one can see positive values for $\dot c$ at the vertical interfaces and negative values at the horizontal interfaces, resulting in N-type rafting. Without dislocations (Fig.2d), the shape of $\dot c$ always stays approximately quadratic, and the values are nearly zero and do not show a sign change. 
\begin{figure}[ht] \centering
	\includegraphics[width=0.95\columnwidth]{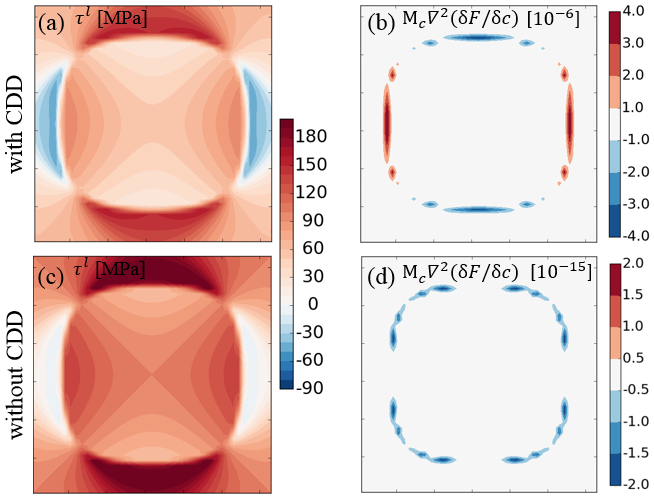}
	\caption{Simulation results at $t=2\times10^4$ \SI{}{s}:  long range shear stress  with CDD (a) and  without CDD (c);  evolution $\dot c$ of $\gamma/\gamma^\prime$ composition field with CDD (b) and  without CDD (d).}
	\label{fig:fig2_v1}
\end{figure}

Fig.3 compares the plastic strain and the plastic strain rate with experimental data at \SI{1253}{K} and \SI{200}{MPa} \cite{Acta_2000_Link}. The plastic strain increases fast in the beginning, which is due to fast motion of dislocations until an increasing number is slowed down within the \ggp\ interface. 'Interface dislocations' only move slowly as the phase boundary evolves. This causes the plastic strain to slowly saturate, as opposed to the experiment where it increases  linearly due to threading dislocations. This discrepancy is due to the fact that  dislocation line length increase is not contained in our idealized model. 
%
The strain rate constantly decreases with time, i.e. strain hardening dominates the early stage of rafting. Thermally activated effects (e.g. dislocation climb) are not considered in the present work. Therefore,  after $10^4$ \SI{}{s} the simulated strain rate is lower than the experimental data. 
\begin{figure}[ht] \centering
	\includegraphics[width=\columnwidth]{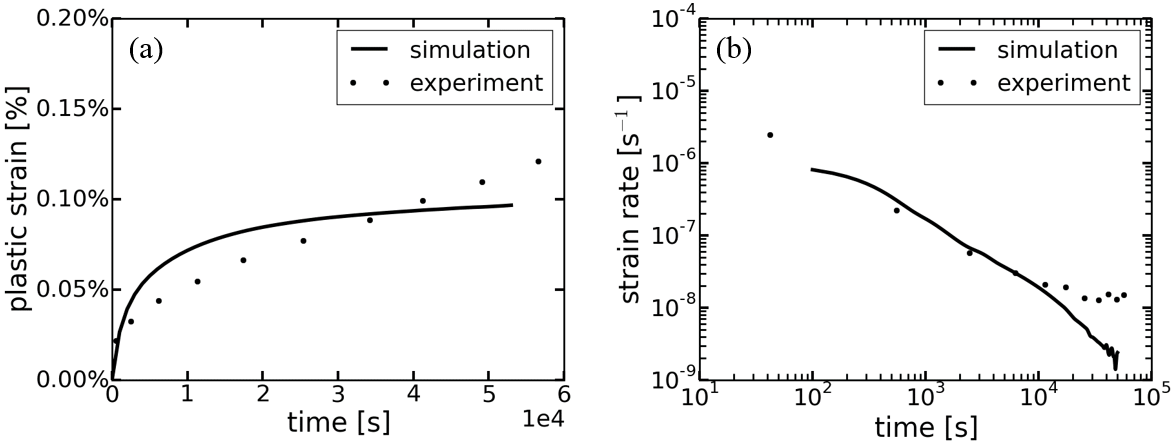}
	\caption{Comparison of simulation and experiment results: (a) plastic strain versus time; (b) strain rate versus time.}
	\label{fig:fig3_v1}
\end{figure}

We presented a minimal model of dislocation-assisted rafting, which can already explain how dislocation and phase microstructure interact during rafting and how this determines the macroscopic creep properties. The ability to represent fluxes of dislocations was identified as one of the \revision{key aspects} of a suitable plasticity model. 
\revision{Nonetheless, this model is just a first step. In order to have a more accurate prediction of microstructure and creep properties, future work will consider a more sophisticated PF model which is able to represent multiple \gp\ variants. In particular when higher plastic strains and creep times will be investigated, climb processes and the associated vacancy diffusion or possibly also dislocations acting as pipe diffusion paths need to be included as well. Concerning the dislocation model, it will be necessary to switch to a more advanced CDD theory (such as the one used in  \cite{2015_MSMSE_Stefan}) which is able also to represent curved dislocations.} 

\section*{Acknowledgment}
Financial support from the Deutsche Forschungsgemeinschaft (DFG) through Research Unit FOR1650 'Dislocation-based Plasticity' (DFG grants SA2292/1-2 and ZA171/7-1 ) is gratefully acknowledged.
\section*{References}

\bibliography{literature}

\begin{thebibliography}{20}
\expandafter\ifx\csname natexlab\endcsname\relax\def\natexlab#1{#1}\fi
\providecommand{\url}[1]{\texttt{#1}}
\providecommand{\href}[2]{#2}
\providecommand{\path}[1]{#1}
\providecommand{\DOIprefix}{doi:}
\providecommand{\ArXivprefix}{arXiv:}
\providecommand{\URLprefix}{URL: }
\providecommand{\Pubmedprefix}{pmid:}
\providecommand{\doi}[1]{\href{http://dx.doi.org/#1}{\path{#1}}}
\providecommand{\Pubmed}[1]{\href{pmid:#1}{\path{#1}}}
\providecommand{\bibinfo}[2]{#2}
\ifx\xfnm\relax \def\xfnm[#1]{\unskip,\space#1}\fi
\bibitem[{Pollock and Argon(1992)}]{Pollock1992_AMM40}
\bibinfo{author}{T.~Pollock}, \bibinfo{author}{A.~Argon},
  \bibinfo{journal}{Acta. Metall. et. Mater.} \bibinfo{volume}{{40}}
  (\bibinfo{year}{1992}) \bibinfo{pages}{{1--30}}.
\bibitem[{Mughrabi and Tetzlaff(2000)}]{Mughrabi2000_AEM2}
\bibinfo{author}{H.~Mughrabi}, \bibinfo{author}{U.~Tetzlaff},
  \bibinfo{journal}{Adv. Eng. Mater.} \bibinfo{volume}{2}
  (\bibinfo{year}{2000}) \bibinfo{pages}{319--326}.
\bibitem[{Reed et~al.(2009)Reed, Tao, and Warnken}]{Reed2009_ActaMater57}
\bibinfo{author}{R.~Reed}, \bibinfo{author}{T.~Tao},
  \bibinfo{author}{N.~Warnken}, \bibinfo{journal}{Acta. Mater.}
  \bibinfo{volume}{57} (\bibinfo{year}{2009}) \bibinfo{pages}{5898--5913}.
\bibitem[{Steinbach(2009)}]{Steinbach2009_MSMSE17}
\bibinfo{author}{I.~Steinbach}, \bibinfo{journal}{Model. Simul. Mater. Sci.
  Eng.} \bibinfo{volume}{17} (\bibinfo{year}{2009}) \bibinfo{pages}{073001}.
\bibitem[{Emmerich(2008)}]{Emmerich2008_AiP57}
\bibinfo{author}{H.~Emmerich}, \bibinfo{journal}{Adv. Phys.}
  \bibinfo{volume}{57} (\bibinfo{year}{2008}) \bibinfo{pages}{1--87}.
\bibitem[{Cottura et~al.(2012)Cottura, Le~Bouar, Finel, Appolaire, Ammar, and
  Forest}]{JMPS_2012_Cottura}
\bibinfo{author}{M.~Cottura}, \bibinfo{author}{Y.~Le~Bouar},
  \bibinfo{author}{A.~Finel}, \bibinfo{author}{B.~Appolaire},
  \bibinfo{author}{K.~Ammar}, \bibinfo{author}{S.~Forest}, \bibinfo{journal}{J.
  Mech. Phys. Solids.} \bibinfo{volume}{60} (\bibinfo{year}{2012})
  \bibinfo{pages}{1243--1256}.
\bibitem[{Gaubert et~al.(2010)Gaubert, Le~Bouar, and Finel}]{PM_2010_Gaubert}
\bibinfo{author}{A.~Gaubert}, \bibinfo{author}{Y.~Le~Bouar},
  \bibinfo{author}{A.~Finel}, \bibinfo{journal}{Philos. Mag.}
  \bibinfo{volume}{90} (\bibinfo{year}{2010}) \bibinfo{pages}{375--404}.
\bibitem[{Finel and Rodney(2000)}]{2000_MRS_Finel}
\bibinfo{author}{A.~Finel}, \bibinfo{author}{D.~Rodney}, in:
  \bibinfo{booktitle}{MRS Fall Meeting, Boston, MA}.
\bibitem[{Wang et~al.(2001)Wang, Jin, Cuitino, and
  Khachaturyan}]{2001_Acta_Khachaturyan}
\bibinfo{author}{Y.~U. Wang}, \bibinfo{author}{Y.~Jin},
  \bibinfo{author}{A.~Cuitino}, \bibinfo{author}{A.~Khachaturyan},
  \bibinfo{journal}{Acta. Mater.} \bibinfo{volume}{49} (\bibinfo{year}{2001})
  \bibinfo{pages}{1847--1857}.
\bibitem[{Rodney et~al.(2003)Rodney, Le~Bouar, and Finel}]{2003_Acta_Finel}
\bibinfo{author}{D.~Rodney}, \bibinfo{author}{Y.~Le~Bouar},
  \bibinfo{author}{A.~Finel}, \bibinfo{journal}{Acta. Mater.}
  \bibinfo{volume}{51} (\bibinfo{year}{2003}) \bibinfo{pages}{17--30}.
\bibitem[{Zhou et~al.(2010)Zhou, Shen, Mills, and Wang}]{PM_2010_Zhou}
\bibinfo{author}{N.~Zhou}, \bibinfo{author}{C.~Shen},
  \bibinfo{author}{M.~Mills}, \bibinfo{author}{Y.~Wang},
  \bibinfo{journal}{Philos. Mag.} \bibinfo{volume}{90} (\bibinfo{year}{2010})
  \bibinfo{pages}{405--436}.
\bibitem[{Groma et~al.(2003)Groma, Csikor, and Zaiser}]{Acta_2003_Michael}
\bibinfo{author}{I.~Groma}, \bibinfo{author}{F.~F. Csikor},
  \bibinfo{author}{M.~Zaiser}, \bibinfo{journal}{Acta. Mater.}
  \bibinfo{volume}{51} (\bibinfo{year}{2003}) \bibinfo{pages}{1271--1281}.
\bibitem[{Sandfeld et~al.(2013)Sandfeld, Monavari, and
  Zaiser}]{MSMSE_2013_Stefan}
\bibinfo{author}{S.~Sandfeld}, \bibinfo{author}{M.~Monavari},
  \bibinfo{author}{M.~Zaiser}, \bibinfo{journal}{Model. Simul. Mater. Sci.
  Eng.} \bibinfo{volume}{21} (\bibinfo{year}{2013}) \bibinfo{pages}{1--22}.
\bibitem[{Kondo et~al.(2014)Kondo, Kubo, Miura, Murata, and
  Yoshinari}]{MATEC_Kondo_2014}
\bibinfo{author}{Y.~Kondo}, \bibinfo{author}{Y.~Kubo},
  \bibinfo{author}{N.~Miura}, \bibinfo{author}{Y.~Murata},
  \bibinfo{author}{A.~Yoshinari}, in: \bibinfo{booktitle}{MATEC Web of
  Conferences}, volume~\bibinfo{volume}{14}, p. \bibinfo{pages}{20003}.
\bibitem[{J{\'a}come et~al.(2013)J{\'a}come, N{\"o}rtersh{\"a}user, Heyer,
  Lahni, Frenzel, Dlouhy, Somsen, and Eggeler}]{Acta_2013_come}
\bibinfo{author}{L.~A. J{\'a}come}, \bibinfo{author}{P.~N{\"o}rtersh{\"a}user},
  \bibinfo{author}{J.-K. Heyer}, \bibinfo{author}{A.~Lahni},
  \bibinfo{author}{J.~Frenzel}, \bibinfo{author}{A.~Dlouhy},
  \bibinfo{author}{C.~Somsen}, \bibinfo{author}{G.~Eggeler},
  \bibinfo{journal}{Acta. Mater.} \bibinfo{volume}{61} (\bibinfo{year}{2013})
  \bibinfo{pages}{2926--2943}.
\bibitem[{Zhou et~al.(2008)Zhou, Shen, Mills, and Wang}]{Zhou_2008_Acta}
\bibinfo{author}{N.~Zhou}, \bibinfo{author}{C.~Shen},
  \bibinfo{author}{M.~Mills}, \bibinfo{author}{Y.~Wang},
  \bibinfo{journal}{Acta. Mater.} \bibinfo{volume}{56} (\bibinfo{year}{2008})
  \bibinfo{pages}{6156--6173}.
\bibitem[{Zhou et~al.(2007)Zhou, Shen, Mills, and Wang}]{Zhou_2007_Acta}
\bibinfo{author}{N.~Zhou}, \bibinfo{author}{C.~Shen},
  \bibinfo{author}{M.~Mills}, \bibinfo{author}{Y.~Wang},
  \bibinfo{journal}{Acta. Mater.} \bibinfo{volume}{55} (\bibinfo{year}{2007})
  \bibinfo{pages}{5369--5381}.
\bibitem[{Miura et~al.(2011)Miura, Kurita, Hayashi, and
  Kondo}]{ISIJ_2011_Miura}
\bibinfo{author}{N.~Miura}, \bibinfo{author}{K.~Kurita},
  \bibinfo{author}{Y.~Hayashi}, \bibinfo{author}{Y.~Kondo},
  \bibinfo{journal}{ISIJ International} \bibinfo{volume}{51}
  (\bibinfo{year}{2011}) \bibinfo{pages}{663--668}.
\bibitem[{Link et~al.(2000)Link, Epishin, Br{\"u}ckner, and
  Portella}]{Acta_2000_Link}
\bibinfo{author}{T.~Link}, \bibinfo{author}{A.~Epishin},
  \bibinfo{author}{U.~Br{\"u}ckner}, \bibinfo{author}{P.~Portella},
  \bibinfo{journal}{Acta. Mater.} \bibinfo{volume}{48} (\bibinfo{year}{2000})
  \bibinfo{pages}{1981--1994}.
\bibitem[{Sandfeld and Po(2015)}]{2015_MSMSE_Stefan}
\bibinfo{author}{S.~Sandfeld}, \bibinfo{author}{G.~Po},
  \bibinfo{journal}{Model. Simul. Mater. Sci. Eng.} \bibinfo{volume}{23}
  (\bibinfo{year}{2015}) \bibinfo{pages}{085003}.

\end{thebibliography}
\end{document}